\pdfoutput=1

\documentclass[11pt]{article}
\usepackage{coling2016}
\usepackage{times}
\usepackage{url}
\usepackage{latexsym}

\usepackage{graphicx}
\usepackage{caption}
\usepackage{subcaption}
\usepackage{changepage}

\usepackage{color}

\usepackage{xspace}

\newcommand{\CMUAff}[0]{\ensuremath{2}\xspace}
\newcommand{\SinicaAff}[0]{\ensuremath{1}\xspace}

\newcommand{\system}{EmotionPush\xspace}

\title{Sensing Emotions in Text Messages:\\An Application and Deployment Study of \system
}

\author{
Shih-Ming Wang$^{\SinicaAff}$~~~
Chun-Hui Li$^{\SinicaAff}$~~~
Yu-Chun Lo$^{\SinicaAff}$\\
\textbf{Ting-Hao (Kenneth) Huang$^{\CMUAff}$~~~
Lun-Wei Ku$^{\SinicaAff}$}\\
$^{\SinicaAff}$~Academia Sinica, Taipei, Taiwan.
\\\{ipod825, iamscli.tw\}@gmail.com, howard.lo@nlplab.cc, lwku@iis.sinica.edu.tw\\
$^{\CMUAff}$~Carnegie Mellon University, Pittsburgh, PA, USA. tinghaoh@cs.cmu.edu
}

\date{August 2016}

\begin{document}
\maketitle


\begin{abstract}

Instant messaging and push notifications play important roles in modern digital life.
To enable robust sense-making and rich context awareness in computer mediated communications, we introduce \emph{\system}, a system that automatically conveys the emotion of received text with a colored push notification on mobile devices. \system is powered by state-of-the-art emotion classifiers
and is deployed for Facebook Messenger clients on Android. The study showed that the system is able to help users prioritize interactions.



\end{abstract}

\section{Introduction}

\begin{figure}[htbp]
    \centering
    \includegraphics[width=0.99\columnwidth]{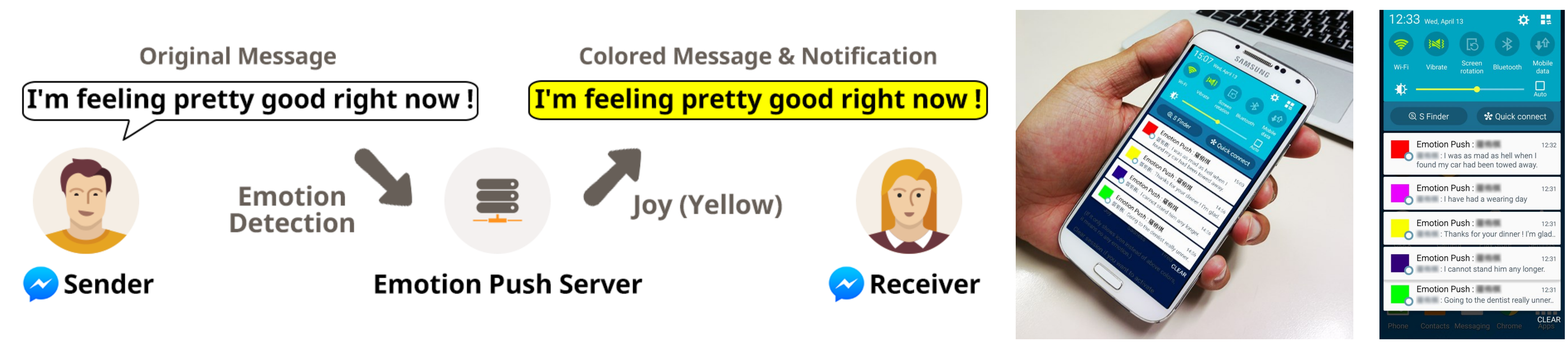}
    \caption{System overview and user scenario of \system. The server detects the emotion of the message and sends a colored notification to indicate the sender's emotion on the receiver's device.}
    \label{fig:system-architecture}
\end{figure}

%
%
\blfootnote{
    %
    %
    %
    %
    %
    
     \hspace{-0.65cm}  
     This work is licensed under a Creative Commons 
     Attribution 4.0 International License.
     License details:
     \url{http://creativecommons.org/licenses/by/4.0/}
}


\noindent Text-based communication plays a big part in computer mediated communications. With the advent of mobile devices, instant messaging and push notifications have become integral to modern digital life. However, text-based chatting is limited in both expressing emotions and building trust amongst users. Text-only chatting has been shown to result in worse communications and trust levels than face-to-face, video, and audio virtual communications~\cite{bos2002effects}. One study also demonstrated that knowing other people's emotions is crucial for collaboration, yet is surprisingly challenging for users via computer mediated communications~\cite{eligio2012emotion}.

In response, previous works have attempted to automatically visualize emotions on text-based interfaces by visualizing the emotion dynamics in a document~\cite{liu2003visualizing},
providing haptic feedback via wearable equipment~\cite{tsetserukou2009ifeel_im}, or changing font sizes according to recognized emotions~\cite{Yeo:CHIEA08}. However, these explorations were primarily developed based on rule-based emotion detectors, which were shown to perform significantly worse than machine-learning algorithms~\cite{wu2006emotion}. On the other hand, some researchers proposed to add new features, such as kinetic typography~\cite{bodine2003kinetic,forlizzi2003kinedit,Lee:2006:UKT:1142405.1142414,lee2002kinetic}, affective buttons~\cite{Broekens:ACII09}, and two-dimensional representations~\cite{Sanchez:CLIHC05,Sanchez:IHC06}, on top of traditional chatting interfaces to allow users express emotions. Others studies have attempted to incorporate the user's body signals, such as fluctuating skin conductivity levels~\cite{dimicco2002conductive}, thermal feedback~\cite{wilsonhot}, or facial expression~\cite{el2004faim}, in instant messaging applications. However, after decades of research, these features are largely absent in modern instant messaging clients.

In this paper, we introduce \textbf{\emph{\system}}, a system that displays colored icons on push notifications (as shown in Figure~\ref{fig:system-architecture}) to signal emotions conveyed in received messages. \system is powered by machine learning technologies with state-of-the-art performances. Built on top of the long-lasting development of emotion detection, we applied the techniques to a real-world chatting environment to determine how well the system works for individual users. Our contribution is two-fold: 1) We created \system, the first system powered by modern machine-learning emotion classification technology to convey emotions for instant messages, and 2) we deployed the system on a widely-used instant messaging client on mobile devices, 
Facebook's Messenger, to examine the feasibility of the emotion feedback.

\section{\system System}

Similar to most mobile apps, \system adopted a client-server architecture (as shown in Figure~\ref{fig:system-architecture}.) When the user (receiver) receives a message via the instant messaging client, our system uses the text of the message to recognize its corresponding emotion, and then notifies the user (receiver) via push notification with a colored icon on his/her mobile device. We developed the \system \emph{client} as an Android application\footnote{\system is available at Google Play: https://play.google.com/store/apps/details?id=tw.edu.sinica.iis.emotionpush}, specifically for Facebook's Messenger (https://www.messenger.com/). The screen shot and user scenario of the app are shown in Figure~\ref{fig:system-architecture}. The \system \emph{server} was implemented as a stand-alone web server powered by pre-trained emotion classification models.

\paragraph{Visualizing Emotions}
\label{sec:visualizing-emotions}

\begin{figure}[t]
    \centering
    \includegraphics[width=0.99\columnwidth]{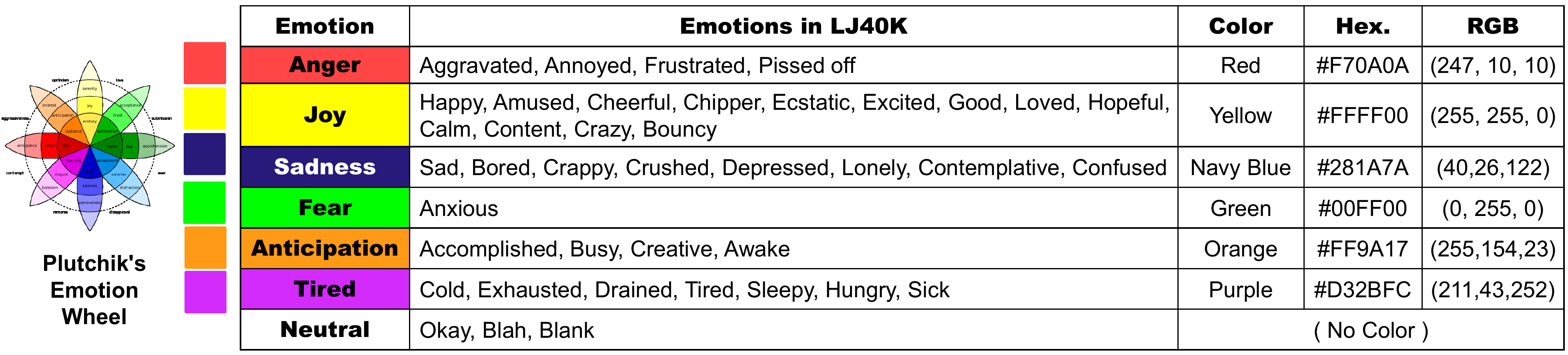}
	\caption{Visualizing emotion colors with Plutchik's Emotion Wheel. The 40 emotion categories of LJ40K are compacted into 7 main categories, each has a corresponding color on the emotion wheel.}
    \label{fig:plutchik}
\end{figure}

\system uses 7 colors to represent 7 emotions, as shown in Figure~\ref{fig:plutchik}. 
This schema was designed as follows: First, we focused on emotions commonly connected with life events, unlike benchmarks such as~\cite{nakov2016semeval} which typically focus on general social media data. To simplify the mapping between emotions and text, we also decided to apply a \emph{categorical representation} (e.g. \textit{Anger}, \textit{Joy}, etc.)~\cite{klein2002computer} of emotions instead of a dimensional representation (valence, arousal)~\cite{Sanchez:IHC06}. Second, we utilized the emotion categories and data provided in \emph{LiveJournal} (http://www.livejournal.com/). LiveJournal is a website where users post what they feel and tag each post with a corresponding emotion. The \emph{LJ40k} corpus~\cite{leshed2006understanding}, a dataset that contains 1,000 blog posts for each of the 40 most common emotions on LiveJournal, was adopted to learn which emotions we should watch for in \system and to train the emotion classifiers. Finally, to reduce users' cognitive load, the original 40 emotions were compacted into 7 main emotions according to Plutchik's Emotion Wheel color theme~\cite{plutchik1980emotion}, as shown in Figure~\ref{fig:plutchik}.

\paragraph{Emotion Classification \& Evaluation}
\label{sec:exp_and_eval}

\system's 7 classifiers were trained on the LJ40k dataset with 7 compacted emotion labels. Each classifier is a binary classifier that indicates if the current message belongs to one of the 7 compacted emotions. The message was inputted into each classifier to obtain the probability of each compacted emotion label. Then the label of the highest probability was selected as the predicted emotion label of the current message.

To compare the performance of our approach with that of previous works on the LJ40k dataset~\cite{yang2013quantitative}, we replicated our classification method and features to predict the original \emph{40} emotions in LJ40k. Forty binary-class classifiers (one emotion each) were developed by using LibSVM~\cite{fan2008liblinear} with a radial basis function kernel. We chose to develop 40 classifiers instead of one 40-class classifier to a) better compare our results to Yang and Liu~\shortcite{yang2013quantitative}, b) achieve better performance, and c) extend \system to a multi-labeling system in the future. To form a balanced training set for each emotion, we randomly selected 800 posts from LJ40K as positive examples and 800 posts of the other 39 emotions as negative examples. Aware of various features proposed for affect~\cite{Ku:2009:UMS:1699648.1699672,ku2011predicting,balahur2014computational,poria2014emosenticspace,tang2014learning,xu2015word}, we used a 300-dimension word vectors trained on Google News~\cite{mikolov2013distributed} (https://code.google.com/archive/p/word2vec/) to represent each post. The model's parameters were tuned via a 10-fold cross-validation process.

\begin{figure}[t]
    \centering
    \includegraphics[width=0.99\columnwidth]{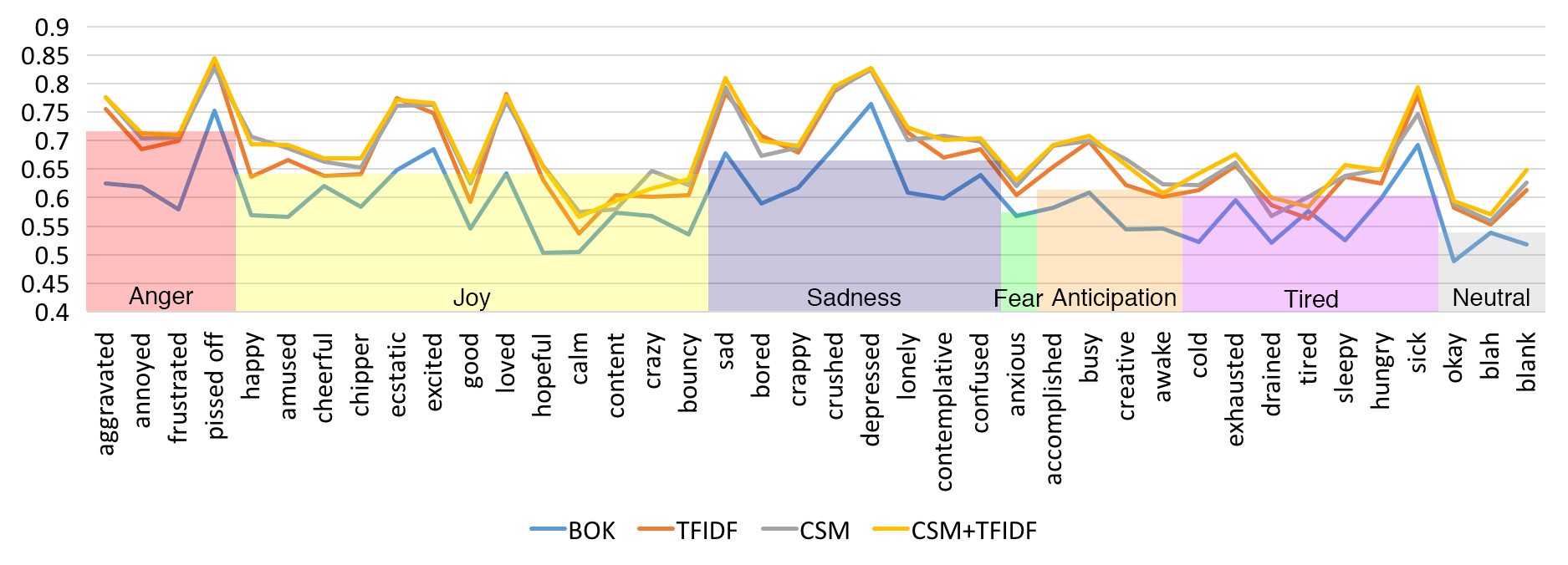}
    \caption{AUC scores of the binary classifiers corresponding to the 40 LiveJournal emotions and the 7 major emotions in \system.}
    \label{fig:40-emotion-compare}
\end{figure}

The evaluations were performed on the held-out testing set that contains 8,000 posts (200 posts for each emotion). The AUC, the area under the receiver operating characteristic curve, was calculated. For the 40 emotions, our classifiers achieved an average AUC of 0.6788, which was comparable to the state-of-the-art performance, 0.6851, reported by~\cite{yang2013quantitative}. Figure~\ref{fig:40-emotion-compare} shows the performance of each emotion using different features.
We observed that classifiers performed worse on \textit{Blank}, \textit{Okay}, \textit{Drained}, and \textit{Sleepy}.
These low-performance emotions can be roughly classified into two categories: 1) \emph{vague emotions}, such as \textit{Blank} and \textit{Okay}, are difficult to model, as people tend to interpret them differently, and 2) \emph{similar emotions}, such as \textit{Drained} and \textit{Sleepy}, tend to overlap significantly, and thus hinder us from distinguishing them.

Furthermore, the colored bars in Figure~\ref{fig:40-emotion-compare} show the classification performances for the 7 compacted major emotions for \system.
The classifiers of compacted emotions \textit{Joy}, \textit{Sadness} and \textit{Anger} performed best among all 7 emotions, while \textit{Neutral} and \textit{Fear} performed worse, which might be because these two compacted emotions were made up of fewer LiveJournal emotions than others, as shown in Figure~\ref{fig:plutchik}.
This not only resulted in the lack of training data, but also brought in errors as these few emotions (including \textit{Anxious}, \textit{Okay}, \textit{Blah}, and \textit{Blank}) performed comparably less satisfactory (see Figure~\ref{fig:40-emotion-compare}).

\section{Deployment Study}



In this study, we aimed to test whether \system can change the priority of interactions in instant messages on mobile devices. Therefore, we deployed \system to Google Play, and recruited 8 native English speakers who frequently used Facebook's Messenger app. The experiment lasted 12 days. We investigated the effect of \system by turning the color feedback off (for the first 5 days, noted as the \emph{first week}) and on (for the latter 7 days, \emph{second week}). We then analyzed changes in participants' priorities of reading and responding to messages.

During the entire study, we collected 6,288 messages in total, 3,844 read counts (first read of a message sequence) and 3,769 response counts (first response). The overall average score obtained from participants was above the average (2.375 over 4) for the question ``\system can predict emotion colors correctly.'' Moreover, participants did not think wrongly predicted emotions would harm their chatting experience (average score 1.375 over 4).



\begin{figure}[t]
    \centering
    \includegraphics[width=0.8\textwidth]{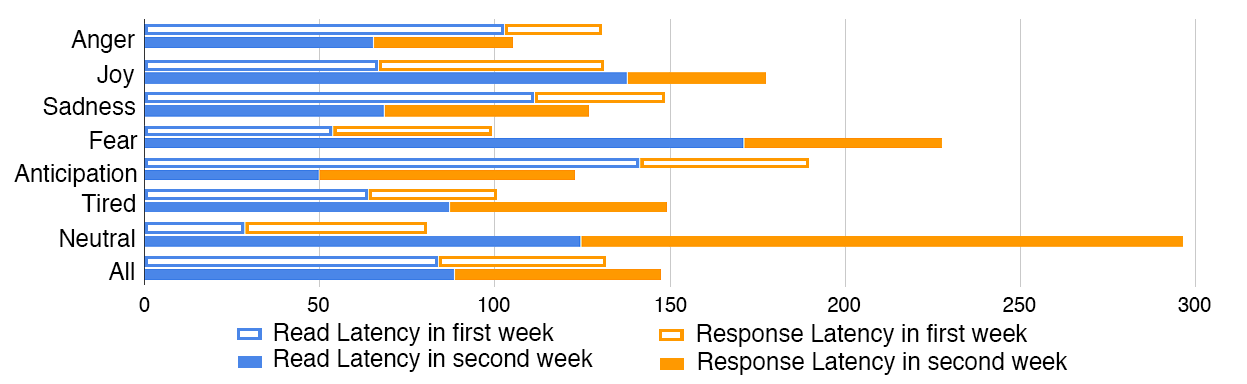}
    \caption{The results of two-week deployment study of \system. This chart shows the users' read and response latencies when color feedback is on (first week) and off (second week), respectively.}
    \label{fig:first_week}

\end{figure}

From this study, we found that \system helps \textbf{prioritize interactions}. The average read latencies and response latencies for the first week and the second week are shown in Figure~\ref{fig:first_week}. We can observe that after the emotion colors were pushed, user's behavior changed accordingly. Especially for \textit{Joy}, \textit{Anger}, and \textit{Sadness}, the priority of the instant message interactions changed in two interesting directions: 1) \textit{Joy} (blue bar: with an increasing read latency from first week to second week) was read more slowly while \textit{Sadness} and \textit{Anger} (blue bar: with a decreasing read latency) was read more quickly, and 2) \textit{Joy} was responded to more quickly (orange bar: with a decreasing response latency) while \textit{Sadness} and \textit{Anger} were responded to more slowly (orange bar: with an increasing response latency). The decreasing read latency of \textit{Anger} ($p=0.0019$) and the increasing response latency of \textit{Sadness} ($p=0.0486$)
are significant. This might reveal that messages that are not urgent (\textit{Joy}) can be put aside and read later, while participants are willing to read urgent messages (\textit{Sadness} and \textit{Anger}) earlier. On the other hand, users could casually respond to less urgent messages (i.e., with little thought), whereas urgent messages require more thought.

An interesting thing to mention is the changes of the response latency. Logically speaking, as participants read messages, they should realize the messages' emotions so that their response latency would not change after pushing emotion colors. However, we observed changes in Figure~\ref{fig:first_week}, and there is even a significant difference of the response latencies for \textit{Sadness}. This suggests that the color feedback is not only notifying users, but also influencing their process of composing a response.

Overall, the feedback for using \system was positive. 78\% of participants thought it is a good idea to add the feature of \system to Facebook's Messenger, and the other 22\% wanted to add this feature eventually but just not immediately to wait for a better user interface and a better prediction performance for some emotion categories.

\section{Conclusion and Future Work}

We introduced \system, a system that automatically recognizes and pushes emotions of instant messages and visualizes them to the end-user, which enables the emotion sensing ability on messages and enriches the information in the communications. We believe this research can help us gain more insight into the effect of reinforcing the emotion sensing of robots.
In the future, we plan to add this emotion sensing function in the message composing process and discuss the quality of conversations.





\bibliographystyle{acl}
\bibliography{coling}

\end{document}